\documentclass{PoS}

\newcommand{\beq}{\begin{equation}}
\newcommand{\eeq}{\end{equation}}
\newcommand{\beqa}{\begin{eqnarray}}
\newcommand{\eeqa}{\end{eqnarray}}
\newcommand{\re}{{\rm Real\/}\,}

\newcommand{\tr}{{\rm tr\/}\,}
\newcommand{\om}{\omega}
\newcommand{\rw}{\rho(\omega)}

\newcommand{\rwt}{\rho_2(\omega)}
\newcommand{\gt}{G_E(\tau)}
\newcommand{\glat}{G^{\rm Lat}_E(\tau)}

\newcommand{\tc}{T_c}
\newcommand{\kt}{\kappa/T^3}
\newcommand{\nt}{N_\tau}

\def \etal{{\sl et al.\/}}
\def \jhep{{\sl J.\ H.\ E.\ P.\ }}

\def \pr{{\sl Phys.\ Rev.\/}}
\def \prl{{\sl Phys.\ Rev.\ Lett.\/}}

\title{An estimate of heavy quark momentum diffusion coefficient in 
gluon plasma}

\ShortTitle{Heavy quark momentum diffusion coefficient}

\author{Debasish Banerjee \\
        Albert Einstein Center, Institute for Theoretical Physics, \\ 
        Bern University, Bern, Switzerland \\
        E-mail: \email{dbanerjee@itp.unibe.ch}}

\author{\speaker{Saumen Datta} and Rajiv Gavai \\
        Tata Institute of Fundamental Research\\
        Mumbai, India. \\
        E-mail: \email{saumen@theory.tifr.res.in,gavai@tifr.res.in}}

\author{Pushan Majumdar \\
        Indian Association for the Cultivation of science \\
        Kolkata, India. \\
        E-mail: \email{tppm@iacs.res.in}}

\abstract{We calculate the momentum diffusion coefficient for heavy quarks in
SU(3) gluon plasma at temperatures 1-2 times the deconfinement
temperature. The momentum diffusion coefficient is extracted from a
Monte Carlo calculation of the correlation function of color electric
fields, in the leading order of expansion in heavy quark mass. 
Systematics of the calculation are examined, and compared with
perturbtion theory and other estimates.}

\FullConference{The 30th International Symposium on Lattice Filed Theory\\
		 June 24-29,2012\\
		 Cairns, australia.}

\begin{document}

\section{Introduction}
\label{sec.intro}
The charm and bottom quarks play a very important role as probes of
the medium created in relativistic heavy ion collision experiments.
Since the masses of both these quarks are considerably larger than the
temperatures estimated to have been reached in RHIC and even in LHC,
one expects these quarks to be produced largely in the early state of
the collision. They therefore allow us to look at the medium at its
early times. Besides, since the nature of energy loss and of collective
behavior of the heavy quarks are different from those of the light
quarks, a study of the heavy quark jets and the flow of the heavy
hadrons leads to crucial insights into the way the medium interacts.

As collision with a thermal quark does not change the energy of a
heavy quark substantially, one would expect the thermalization time of
the heavy quark to be large. Since elliptic flow develops early, it is
reasonable to expect that the elliptic flow will show a mass
hierarchy: $v_2^h \gg v_2^D \gg v_2^B$, where D and B refer to generic
mesons in D (one valence charm) and B (one valence bottom) family, and
$h$ refers to the light hadrons. Similarly, the nuclear suppression
factor can be intuitively expected to be closer to 1 for the heavy-light
mesons: $R^h_{AA} \ll R^D_{AA} \ll R^B_{AA}$.

Since the typical energy loss in a hard collision with the thermal
particles is $\sim T$, for a thermal heavy quark with $M \gg T, \ p
\sim \sqrt{M T}$, it takes a large number of collisions for the heavy
quark to change its momentum by $\mathcal{O}(1)$. Therefore, one can
picture scattering with thermal quarks as uncorrelated momentum kicks,
and use a Langevin description for the motion of the heavy quark in
the thermal medium \cite{svetitsky,moore}: 
\beq 
\frac{d p_i}{dt} \ = \ - \eta_D p_i \ +
\ \xi_i(t), \qquad \langle \xi_i(t) \xi_j(t^\prime) \rangle \ = \ \kappa
\ \delta_{i j} \ \delta(t-t^\prime). 
\label{eq.langevin} \eeq
The momentum diffusion coefficient,
$\kappa$, is related to the correlation of the force term:
\\ 
\beq 
\kappa \ = \ \frac{1}{3} \ \int_{- \infty}^\infty dt \ \sum_i
\langle F(t) F(0) \rangle .
\label{eq.force} \eeq
$\kappa$ has been determined in perturbation theory\cite{moore}.
The fluctuation-dissipation theorem relates $\eta_D$ and $\kappa$
\cite{kapusta}: \\ 
\beq 
\eta_D = \frac{\kappa}{2 M T}.
\label{eq.fd} \eeq

The heavy quark elliptic flow, and $R_{AA}$, has been measured in RHIC
\cite{rhic}.  While a Langevin formalism based description,
Eq. (\ref{eq.langevin}), seems to describe the $p_T$ dependence of the
heavy quark flow quite well, the value of $\kappa$ needed to describe
the data \cite{rhic} is much larger than the leading order (LO)
perturbation theory prediction. Recently, the next-to-leading order
(NLO) calculation of $\kappa$ has been performed in the static quark
limit \cite{cm}. While the NLO result is, encouragingly, much larger,
the large change between LO and NLO also points to the unreliability
of perturbation theory in the temperature regime of a few times the
transition temperature $T_c$. A non-perturbative evaluation, if
possible, will greatly add to our understanding of the medium response
to a heavy quark probe.

Here we report the results of an non-perturbative estimation of
$\kappa$ using lattice QCD in the quenched approximation (i.e., for a
gluon plasma). In the next section we outline the methodology, and in
the following section we discuss the results. A more detailed
discussion, including examination of various systematic errors, can be
found in Ref. \cite{prd}.

\section{Calculational Details:}
The calculation of the heavy quark momentum diffusion coefficient,
$\kappa$, is a nontrivial problem for lattice QCD. On the lattice, one
only calculates the Euclidean-time Matsubara correlators, while
$\kappa$, and other transport coefficients, are directly connected to
the real-time retarded correlators of suitable currents \cite{meyer}.
An analytical continuation of the Matsubara correlator to real time is
required to extract $\kappa$ from it. Such a continuation of numerical
data is very difficult (see \cite{meyer} for a recent
review). On top of that, $\kappa$ is related to the width of a narrow
peak in the spectral function of the quark number current, making it
very difficult to extract it reliably.

A strategy of extracting $\kappa$ for an infinitely heavy (``static'')
quark has been formulated in Ref. \cite{ct,clm} which alleviates this
second problem somewhat. In the static limit, the propagation of heavy
quarks is replaced by Wilson lines, and the correlator of
Eq. (\ref{eq.force}) reduces to the evaluation of retarded correlator
of electric fields connected by Wilson lines \cite{ct}.  This
correlator can be analytically continued to Euclidean time \cite{clm}.
The lattice discretization of the Euclidean correlator leads to \beq
\glat \ = \ -\frac{1}{3 L} \ \sum_{i=1}^3 \ \left\langle \re \ \tr
\ \left[ U(\beta, \tau) \ E_i(\tau, \vec{0}) \ U(\tau,0) \ E_i(0,
  \vec{0}) \right] \right\rangle ,
\label{eq.cor} \eeq
where $U(\tau_1, \tau_2)$, the timelike gauge connection between
points $(\vec{x}, \tau_1)$ and $(\vec{x}, \tau_2)$, is the phase
factor associated with the evolution of an infinitely heavy quark in
imaginary time. $E_i(\tau, \vec{x})$ is the color electric field at
point $(\vec{x}, \tau)$ and $L = \tr U(\beta, 0)$ is the Polyakov
loop.

In order to calculate $\glat$, we use the discretization of the
electric field \cite{clm} \beq E_i (\vec{x}, \tau) \ = \ U_i (\vec{x},
\tau) \ U_4 (\vec{x}+\hat{i}, \tau) \ - \ U_4 (\vec{x}, \tau) \ U_i
(\vec{x} + \hat{4})
\label{eq.E} \eeq
which has good ultraviolet properties. We calculated
$G^{\rm Lat}_E(\tau)$ on a set of SU(3) pure gauge lattices in the
temperature range $1-2 \tc$. It is imperative to use sufficiently fine
lattice spacings, $a$, so that we get a large number of points, $\nt$,
in the $\tau$ direction. We could reliably extract $\kappa$ only from 
lattices with $\nt = 20$ or more. The different lattices for which we will
be quoting estimates of $\kappa$ are shown in Table
\ref{tbl.lattices}.

\begin{table}[hbt]
\begin{center}
\begin{tabular}{ccc|cc}
\hline
$\beta$ & $\nt$ & $T/\tc$ & \# sublattices & \# sublattice updates \\

\hline
6.76  & 20 & 1.04 &  5 & 4000 \\
6.80  & 20 & 1.09 &  5 & 3000 \\
6.90  & 20 & 1.24 &  5 & 2000  \\
7.192 & 24 & 1.50 &  4 & 2000  \\
7.255 & 20 & 1.96 &  5 & 2000  \\
\hline
\end{tabular}
\caption{List of lattices which were used to calculate $\kappa$ from $\gt$. 
Also given are the parameters used for multilevel update: the number of 
sublattices the $\tau$ direction was divided in, and the number of sublattice
averagings before a full lattice update.
\label{tbl.lattices}}
\end{center} \end{table}

A precise calculation of $\glat$ on lattices with such large $\nt$ is
known to be very difficult with a naive updating algorithm. We adapted
the multilevel algorithm \cite{multilevel}, which was devised
precisely for such problems, to calculate Eq. (\ref{eq.cor}). The
lattice is divided into several sublattices. The expectation value of
the correlation functions are first calculated in each sublattice by
averaging over a large number of sweeps in that sublattice while
keeping the boundary fixed.  A single measurement is obtained by
multiplying the intermediate expectation values appropriately. The
number of sublattices and the number of sublattice averagings were
tuned for the various sets, so as to get correlators with a few per cent level
accuracy. The parameters used for the lattices in Table
\ref{tbl.lattices} are also shown in the table. The use of the
multilevel algorithm turned out to be absolutely essential for our
calculation: for large $\tau$, it was up to $\mathcal{O}(10^3)$
times more efficient than a naive updating algorithm. For details
about implementation of the algorithm, and its performance, see Ref. \cite{prd}.

Before extracting $\kappa$, we need to convert $\glat$ to the physical
correlator of the electric field, \beq \gt = Z(a) G^{\rm Lat}_E(\tau)
\label{eq.z} 
\eeq where $Z(a) = Z_E^2$ is the lattice spacing dependent
renormalization factor for the electric field correlator. We use the
tree-level tadpole factor for $Z(a)$ \cite{lm}.  A nonperturbative
evaluation of $Z(a)$ is planned for the future.

\section{Results:}
Standard fluctuation-dissipation relations \cite{kapusta} connect the
momentum diffusion coefficient, $\kappa$, to the low-$\om$ behavior of
the spectral function: \beqa \gt \ &=& \ \int_0^\infty \frac{d
  \om}{\pi} \ \rw \ \frac{\cosh \om(\tau-1/2T)}{\sinh
  \frac{\om}{2T}} \label{eq.spectral} \\ \kappa \ &=& \lim_{\om \to 0}
\frac{2T}{\om} \ \rw .
\label{eq.kp} 
\eeqa

The nontrivial part of the calculation is to extract $\rw$
from $\gt$. In this work we assumed a functional form for $\rw$, so
that calculation of $\rw$ and $\kappa$ become a fitting problem. We
use the simple fit form \beq \rw \ = \ \frac{\kappa}{2 T} \om
\ \Theta(\Lambda - \omega) \ + \ b \om^3,
\label{eq.w}
\eeq 
where the first term in the r.h.s. of Eq. (\ref{eq.w}) is the low-$\om$ 
diffusion part, and $\Lambda$ is an infrared cutoff. As elaborated later, 
we fix $\Lambda=3T$ for our central results and fit for $\kappa, b$. 
For the fit, $\chi^2$ minimization was carried out
with the full covariance matrix included in the definition of
$\chi^2$.  We typically obtained acceptable fits to the correlators
for $\tau a$ in the range $[N_t/4, N_t/2]$, with $\chi^2/{\rm d.o.f}
\sim 1$.  At shorter distances, lattice artifacts start contributing
and the simple form of Eq. (\ref{eq.w}) does not work well. Also using
the leading order lattice correlator instead of the continuum form did
not improve the quality of the fit. We, therefore, restrict ourselves
to the long distance part of the correlator.

\begin{figure}
\centerline{\includegraphics[width=.5\textwidth]
{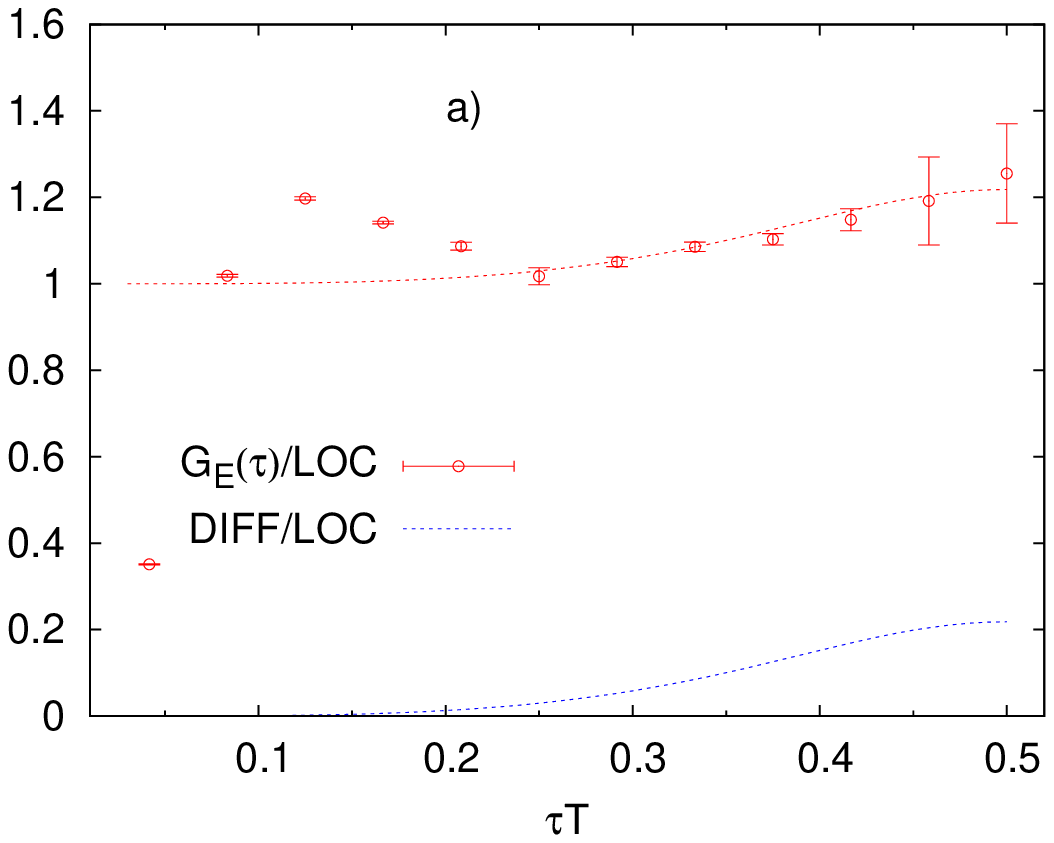}
\includegraphics[width=.5\textwidth]{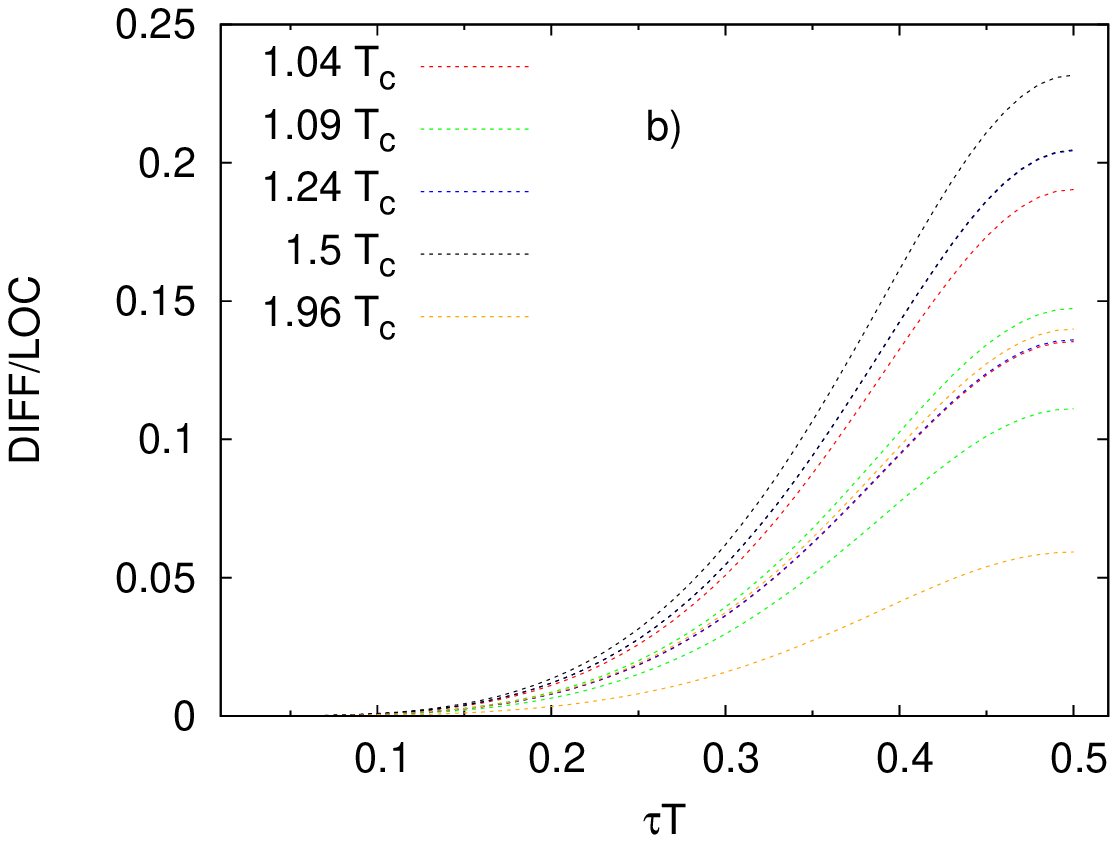}}
\caption{(a) $\gt$ and the correlator reconstructed from the diffusive 
part(DIFF), shown normalized to that reconstructed from the leading order 
part (LOC), at 1.5 $\tc$ ($\beta$ = 7.192); see text. 
(b) The contribution of the diffusive part, normalized by the 
leading order part, at the different temperatures in Table 
\protect \ref{tbl.lattices}.}
\label{fig.diff}
\end{figure}

In order to get a feel for the contribution of the diffusive part of
the spectral function to the correlator, in Fig. \ref{fig.diff} a) we
show the correlators reconstructed from different parts of $\rw$
separately. In this figure we take the $\nt = 24, 1.5 \tc$ data set,
and use the best fit form of Eq. (\ref{eq.w}) (for $\Lambda = 3
T$). The contributions to the total correlation function from the
$\om^3$ part of $\rw$ and that from the diffusive part are calculated
separately using Eq. (\ref{eq.spectral}). In the figure these two
parts are referred to as LOC and DIFF, respectively, and the total
correlator and DIFF are shown normalized by LOC.  While the correlator
is dominated by the contribution from the $b \om^3$ term, the
diffusion term has a substantial contribution near the center of the
lattice.  In Fig. \ref{fig.diff} a) it contributes $\sim$ 17 \% at
$\tau T$ = 0.5.

In Fig. \ref{fig.diff} b) we show the ratio of the diffusive and the
leading order parts of the correlator with varying $\tau$ at the
different temperatures studied in Table \ref{tbl.lattices}. Here we
show the 1-$\sigma$ bands and not the best fit values. At all
temperatures, except the one at the highest temperature, the diffusive
part is seen to reach about 5 \% level by $\tau T \sim 0.3$. Note that
the accuracy of our correlator is better than this. Also no
significant trend of temperature dependence is seen in this figure,
indicating the lack of a strong temperature dependence of $\kappa$ in
this temperature range.

\begin{figure}
\centerline{\includegraphics[width=.5\textwidth]{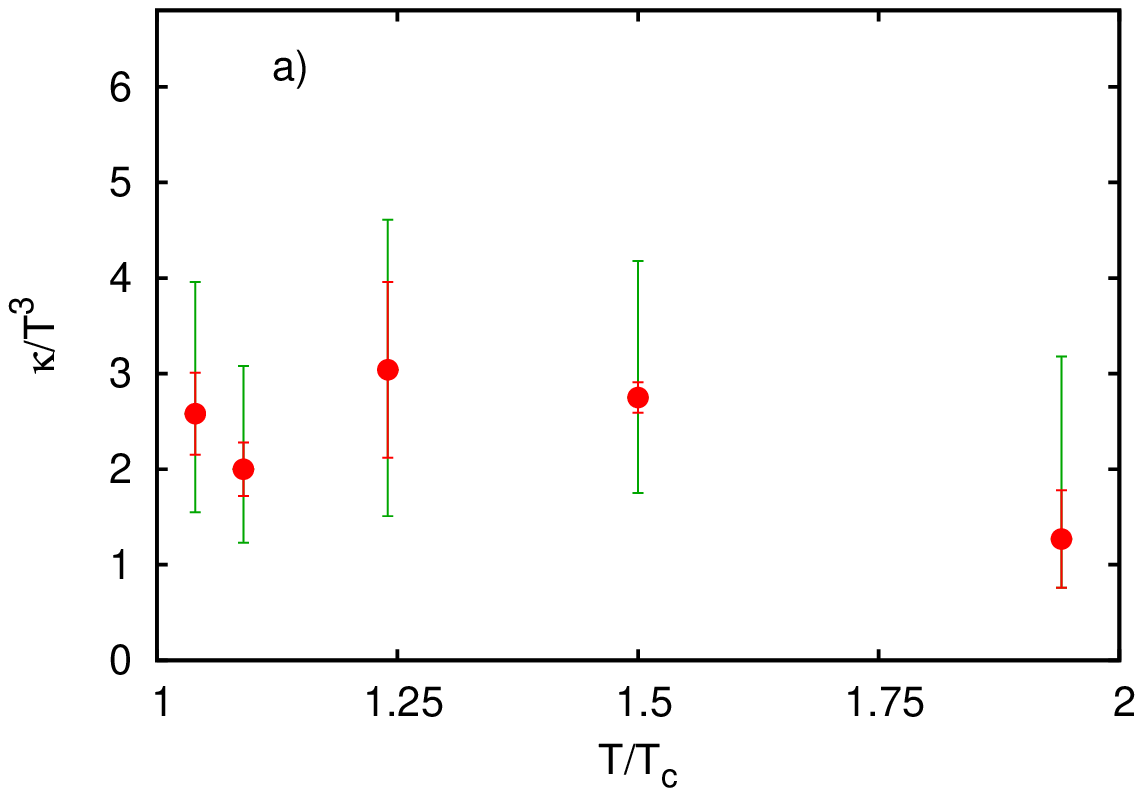}
\includegraphics[width=.5\textwidth]{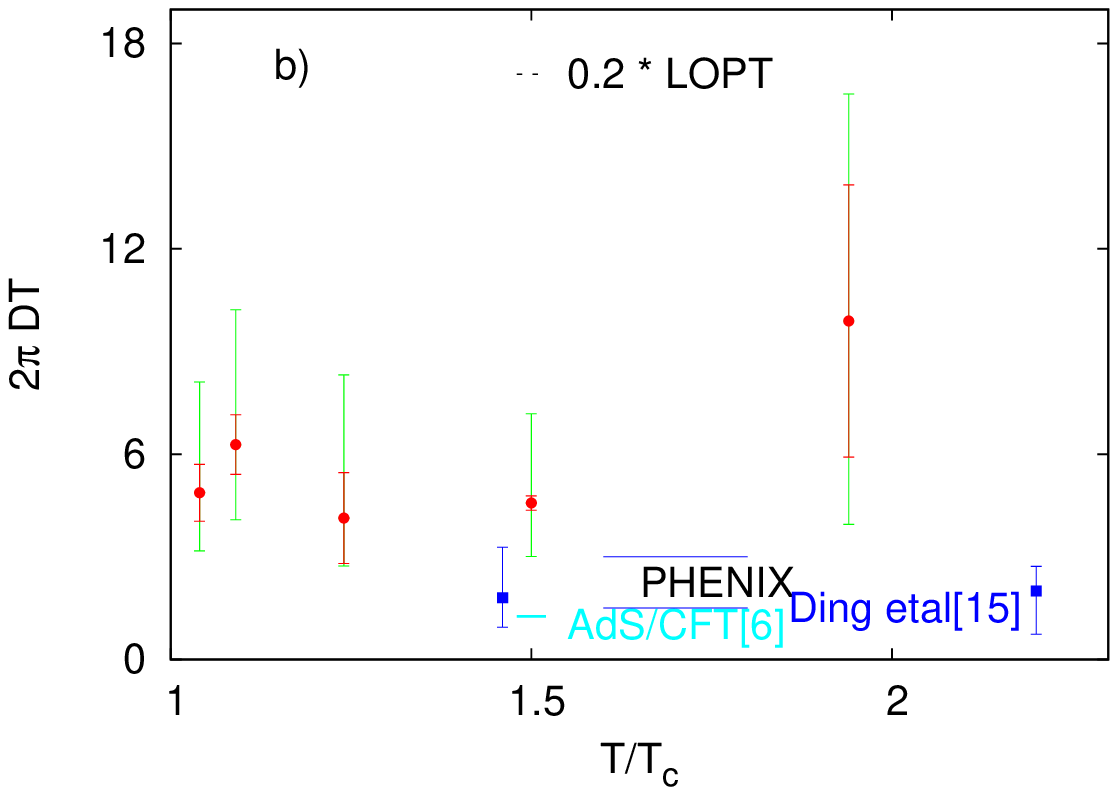}}
\caption{(a) $\kt$ extracted from $\gt$, as explained in the text,
at various temperatures in the gluon plasma. (b) The diffusion 
coefficient, $D$, obtained from $\kappa$ using Eq. (\protect \ref{eq.dt}). 
Also shown are the LO perturbation theory estimate \cite{moore} and 
a different lattice estimate \cite{ding}, both for gluon plasma. 
The experimental value quoted by PHENIX, and an estimate from the
$\mathcal{N}=4$ supersymmetric Yang-Mills theory are also shown.}
\label{fig.result}
\end{figure}

The value of the momentum diffusion coefficient, $\kappa$, obtained
from the analysis outlined above, is shown in Fig. \ref{fig.result}
a). The central error bar corresponds to the purely statistical error,
obtained from a Jackknife analysis. The larger error bar corresponds
to uncertainties due to various systematics:

\begin{itemize}
\item As mentioned above, for the central value shown in the figure,
  we have set $\Lambda = 3 T$. In order to look for the dependence of
  the result on this choice, we have varied $\Lambda$ in the range $[
    2 T, \infty )$. A systematic error band which envelops the central
    values of the fits with varying $\Lambda$ is introduced.

\item The fit form, Eq. (\ref{eq.w}), is a simple model,
  taking into account the leading order perturbative behavior and the
  fact that at small $\omega$, $\rw \propto \om$. In order to test the
  dependence of the extracted value of $\kappa$ on Eq. (\ref{eq.w}) we
  also use an alternate form for the infrared part of $\rw$: \beq \rwt
  \; = \; \kappa \: \tanh \frac{\om}{2 T} \; \Theta(\om - \Lambda) \;
  + \; b \om^3.
\label{eq.t} 
\eeq 
This form is motivated by classical lattice
gauge theory \cite{classical}. The difference between $\kappa$ obtained
from uses of Eq. (\ref{eq.w}) and Eq. (\ref{eq.t}) is also included in
the systematic error.

\item
For the central values in Fig. \ref{fig.result}, we have used the
range $\tau_{\rm min}$ to $\nt /2$, where $\tau_{\rm min}$ is the
smallest $\tau$ for which we got a good $\chi^2$. We looked for the
stability of the fit values for variation of $\tau_{\rm min}$. In all
sets except one, we could get stable fits, with good $\chi^2$ for
$\tau_{\rm min} \ge \nt/4$. For the set at 1.96 $\tc$, where the fit
value was not so stable, we included the variation of the fit value
into our systematic error estimation.

\end{itemize}

A detailed discussion of the sizes of the various systematics can be
found in Ref. \cite{prd}.  Within the small variation of $LT$ our
resources permitted, we did not find a finite volume dependence above
our other errors for $LT \ge 2$.

The value of $\kt$ at 1.5 $\tc$, shown in Fig. \ref{fig.result} a),
agrees within errors with a similar calculation by Francis et
al.  \cite{francis}, while an earlier calculation \cite{meyer2} found
smaller values. It is also an order-of-magnitude larger than the
LO perturbation theory value that can be extracted from
Ref. \cite{moore}.

The experimental results for heavy quark diffusion are usually
presented in terms of the diffusion coefficient $D$, which controls
diffusion of the heavy quark in position space. The Einstein relation,
\begin{equation}
 D = \frac{T}{M \eta_D} = \frac{2 T^2}{\kappa},
\label{eq.dt}
\end{equation}
connects $D$ with $\kappa$. $D$ obtained from Eq. (\ref{eq.dt}) and
Fig. \ref{fig.result} a) is shown in Fig. \ref{fig.result} b). The leading 
order perturbation theory value \cite{moore} is also shown there.

A more direct approach to calculating $D$ from lattice QCD would be to
look at the correlation function of the heavy quark number current,
$\bar{Q} \gamma_i Q$. A calculation of $D$ for the gluon plasma using
such an approach has been presented in Ref. \cite{ding}. Their results are 
also shown in Fig. \ref{fig.result} b), where we have added there statistical
and systematic errors in quadrature. The results are even further from the
perturbation theory estimate and systematically lower than our results, 
although consistent within the large error bars.

Our results are for a gluon plasma, so a comparison with the
experimental results \cite{rhic} requires due caution. At the least, a
comparison of the results in Fig. \ref{fig.result} with the
perturbative results for quenched QCD gives us an indication of how
much the nonperturbative results can change from the perturbative
results in the deconfined plasma at moderate temperatures $< 2
\tc$. Even then, the results are most encouraging since they indicate
that the nonperturbative estimate for $DT$ can easily be an order of
magnitude lower than perturbation theory, bringing it in the right
ballpark to explain the $v_2$ data.

Since dimensionless ratios of various quantities are known to scale
nicely between quenched and full QCD if plotted as function of
$T/T_c$, one can, more optimistically, hope that our results, as plotted in
Fig. \ref{fig.result}, will be even quantitatively close to 
the full QCD values. In this spirit, in Fig. \ref{fig.result} b) we also compare
the lattice results with the experimental data.  The lattice results
seem to be a little above the best range for description of the PHENIX
data \cite{rhic} using the Langevin approach \cite{moore}, though
reasonably close within our large systematics. Interestingly, our
lattice results seem to show very little temperature dependence in the
temperature regime studied here.

As mentioned in Sec. \ref{sec.intro}, the heavy quark diffusion
coefficient has also been calculated in a very different theory, the
$\mathcal{N} = 4$ supersymmetric Yang-Mills theory with the number of
colors, $N_c \to \infty$, at large 't Hooft coupling $\lambda_{tH} =
g^2 N_c$, using AdS/CFT correspondence \cite{ct}.  Of course, this
theory is very different from QCD in many respects.  Moreover, it
crucially exploits symmetries which QCD does not have.  However, to
get a feel for what kind of values of $D$ such a theory would predict for
parameters relevant for QCD at $\sim 1.5 \tc$, we naively set $N_c=3$ and
$\alpha_S$ = 0.23 in the results of Ref. \cite{ct}.  This
gives $DT \simeq 0.2$ (Fig. \ref{fig.result} b), which is lower than,
but in the same ballpark as our estimate.


\begin{thebibliography}{99}
\bibitem{svetitsky}
   B. Svetitsky, \pr D 37 (1988) 2484.
\bibitem{moore}
   G. D. Moore and D. Teaney, \pr C 71 (2005) 064904. 
\bibitem{kapusta}
   J. Kapusta and C. Gale, {\em Finite Temperature Field Theory},
   Cambridge University Press.
\bibitem{rhic}
   A. Adare \etal. (PHENIX Collab.), \prl 98 (2007) 172301. \pr 
C 84 (2011) 044905. \\
   B.I. Abelev \etal. (STAR Collab.), \prl 98 (2007) 192301.
\bibitem{cm} S. Caron-Huot and G. Moore, \jhep 0802 (2008) 081.
\bibitem{prd} D. Banerjee, S. Datta, R. Gavai and P. Majumdar, \pr D 85
(2012) 014510.
\bibitem{meyer} H. B. Meyer, Eur.Phys.J. A47 (2011) 86.
\bibitem{ct}
   J. Casalderrey-Solana and D. Teaney, \pr D 74 (2006) 085012.
\bibitem{clm}
   S. Caron-Huot, M. Laine and G. D. Moore, \jhep 0904 (2009) 053.
\bibitem{multilevel}M. L\"uscher and P. Weisz, \jhep 0109 (2001) 010, 
  \jhep 0207 (2002) 049. 
\bibitem{lm} G. P. Lepage and P. Mackenzie, \pr D 48 (1993) 2250. 
\bibitem{classical} M. Laine, G. Moore, O. Philipsen and M. Tassler, 
\jhep 0905 (2009) 014.
\bibitem{francis} A. Francis, O. Kaczmarek, M. Laine and J. Langelage, 
PoS LATTICE2011 (2011) 202. 
\bibitem{meyer2} H. B. Meyer, New J. Phys. 13 (2011) 035008.
\bibitem{ding} H. T. Ding \etal., \pr D 86 (2012) 014509.
\end{thebibliography}
\end{document}